\documentclass[aps,twocolumn,showpacs,amsmath,amssymb,prl]{revtex4}
\usepackage[dvips]{graphicx}% Include figure files
\usepackage{dcolumn}% Align table columns on decimal point
\usepackage{bm}% bold math

\begin{document}

\title{Correlation Assisted Phonon Softenings and 
the Mott-Peierls Transition in VO$_{2}$}

\author{Sooran Kim, Kyoo Kim, Chang-Jong Kang,}
\author{B. I. Min}
%\email[mail to : ]{bimin@postech.ac.kr}
\affiliation{Department of physics, PCTP,
Pohang University of Science and Technology, Pohang, 790-784, Korea}

\begin{abstract}

To explore the driving mechanisms of the metal-insulator transition (MIT) 
and the structural transition in VO$_{2}$,
we have investigated phonon dispersions of rutile VO$_{2}$ 
(\textit{R}-VO$_{2}$) 
in the DFT and the DFT+$U$ ($U$: Coulomb correlation)
band calculations. 
We have found that the phonon softening instabilities occur in both cases,
but the softened phonon mode only in the DFT+$U$
describes properly both the MIT and the structural transition 
from \textit{R}-VO$_{2}$ to monoclinic VO$_{2}$ (\textit{M$_{1}$}-VO$_{2}$).
This feature demonstrates that the Coulomb correlation effect plays 
an essential role of assisting the Peierls transition in \textit{R}-VO$_{2}$.
We have also found from the phonon dispersion of 
\textit{M$_{1}$}-VO$_{2}$ that \textit{M$_{1}$} structure becomes unstable
under high pressure.
We have predicted a new phase of VO$_{2}$ at high pressure
that has a monoclinic CaCl$_{2}$-type structure with metallic nature.

\end{abstract}


\maketitle

Vanadium dioxide (VO$_{2}$) is one of the most explored 
transition metal oxides
due to its intriguing metal-insulator transition (MIT)
and the concomitant structural transition upon cooling.
At ambient pressure and high temperature, 
VO$_{2}$ has a tetragonal rutile-type structure (\textit{R}-VO$_{2}$)
with metallic nature. Upon cooling below 340 K, \textit{R}-VO$_{2}$
undergoes the structural transition to a monoclinic structure
(\textit{M$_{1}$}-VO$_{2}$) with insulating nature.\cite{Morin59,Goodenough60}
The mechanism of MIT in VO$_{2}$ has been a longstanding subject of controversy.
In the structural transition 
from \textit{R}-VO$_{2}$ to \textit{M$_{1}$}-VO$_{2}$,
V ions construct the dimerization and the zigzag distortion. 
In \textit{R}-VO$_{2}$, V ions are centered at the distorted O$_6$
octahedra, which are edge-shared along the $c$-axis.
Due to the crystal field, V $3d$ states are split into $a_{1g}$ 
(\textit{d$_{\Vert}$}), $e_g^{\pi}$ ($\pi^{*}$), and $e_g^{\sigma}$ 
states in order of energy, and so one electron of V$^{4+}$ ion occupies 
the lowest $a_{1g}$ state. 
While the zigzag distortion increases the energy of $e_g^{\pi}$ bands, 
the dimerization of V-V causes the splitting of $a_{1g}$ 
bands to open the gap at the Fermi level ($E_F$).\cite{Eyert02} 
This kind of structural distortion is explained by a
typical Peierls transition. 

However, the density functional theory (DFT) band approach fails to describe
the insulating nature of \textit{M$_{1}$}-VO$_{2}$ 
properly.\cite{Eyert02,Wentzcovitch94,Liebsch05}
The energy gap at $E_F$ can be obtained only when the extra
Coulomb correlation $U$ effect of V $3d$ electrons is considered,
which indicates that \textit{M$_{1}$}-VO$_{2}$ is a Mott-Hubbard type insulator.
Hence the Mott-Hubbard transition was proposed 
as the mechanism of MIT in VO$_{2}$.\cite{Zylbersztejn75,Rice94,Htkim06}
Beyond the DFT band approach, 
the GW\cite{Continenza99,Gatti07,Sakuma08} or 
the hybrid functional band method\cite{Eyert11}
was employed to describe the insulating nature 
of \textit{M$_{1}$}-VO$_{2}$ properly.
There were also the DMFT (dynamical mean-field theory) approaches 
to explain the insulating nature of \textit{M$_{1}$}-VO$_{2}$ by 
considering the Mott-Hubbard $U$ 
explicitly.\cite{Biermann05,Liebsch05,Lazarovits10,Belozerov12} 
Therefore the consensus at the moment is that some amount
of $U$ is necessary to describe the insulating nature 
of \textit{M$_{1}$}-VO$_{2}$.\cite{Haverkort05,Weber12}
However, whether \textit{R}-VO$_{2}$ is a strongly correlated system
or whether $U$ is necessary for the MIT has not been clarified yet,
despite a few existing studies using the DFT+$U$\cite{Korotin02,Liu10} 
and the DMFT.\cite{Liebsch05,Lazarovits10,Belozerov11}

Besides \textit{R}-VO$_{2}$ and \textit{M$_{1}$}-VO$_{2}$,
other structural types of VO$_{2}$ were reported,
such as monoclinic \textit{M$_{2}$}-VO$_{2}$, \textit{M$_{3}$}-VO$_{2}$ 
and triclinic \textit{T}-VO$_{2}$, which are stabilized
under the uniaxial stress or with doping of 
Cr or Al.\cite{Marezio72,Pouget74,Pouget75,Ghedira77,Ghedira2} 
All the monoclinic phases of \textit{M$_{1}$},
\textit{M$_{2}$}, and \textit{M$_{3}$}-VO$_{2}$ are insulators
at ambient pressure, but at the pressure above ~10 GPa, 
\textit{M$_{1}$}-VO$_{2}$, Cr-doped 
\textit{M$_{2}$}-VO$_{2}$, and Cr-doped \textit{M$_{3}$}-VO$_{2}$
become metallic with some type of monoclinic structure,
so called, \textit{M$_{x}$} phase.\cite{Arcangeletti07,Marini08,Mitrano12}
Still the real structure of \textit{M$_{x}$} phase has not been identified yet.

As described above, there have been extensive electronic structure studies 
on VO$_{2}$. By contrast, there have been only several phonon studies on 
VO$_{2}$.\cite{Srivastava71,Hearnt72,Gervais85,Schilbe02,Schilbe04} 
Especially, there has been neither experimental nor theoretical report 
on the phonon dispersion curve for VO$_{2}$ yet. 
Since the Peierls transition is closely related to the phonon softening
instability, the study of phonon dispersions of VO$_{2}$ is expected to
give a clue to understanding the mechanism of MIT 
in VO$_{2}$.\cite{Brews70,Hearnt72,McWhan74,Terauchi78,Gervais85} 

In this letter, we have revisited the MIT and the structural transition
of VO$_{2}$ by investigating the phonon dispersions of \textit{R}-VO$_{2}$ 
and \textit{M$_{1}$}-VO$_{2}$.
We have found that
\textit{R}-VO$_{2}$ is a strongly correlated system with $U \geq 4.0$ eV,
and the Coulomb correlation effect plays
an essential role in the structural transition 
from \textit{R}-VO$_{2}$ to \textit{M$_{1}$}-VO$_{2}$.
We have also studied the structural stability of \textit{M$_{1}$}-VO$_{2}$
under pressure, and found that \textit{M$_{1}$} phase becomes unstable 
to a phase that seems to be related to \textit{M$_{x}$} phase.

We consider the two phase of VO$_{2}$, \textit{R}-VO$_{2}$ and 
\textit{M$_{1}$}-VO$_{2}$, the space groups of which correspond
to \emph{P}4$_{2}$/\emph{mnm} and \emph{P}2$_{1}$/\emph{c}, 
respectively.\cite{Eyert02}  
For the electronic structure and phonon dispersion calculations,
the pseudo-potential band method and the supercell approach 
that are implemented in VASP\cite{Kresse96} and PHONOPY\cite{Togo08} 
are used. The utilized exchange-correlation functional is the 
generalized gradient approximation (GGA). 
The adopted values of $U$ and $J$ in the DFT+$U$ are 4.2 eV and 0.8 eV, 
respectively.\cite{Liebsch05} 
All the phonon calculations were done after the full 
relaxation of the volume and atomic positions.\cite{Comp}
The initial structural parameters of \textit{R}-VO$_{2}$ and 
\textit{M$_{1}$}-VO$_{2}$ before the relaxation are taken 
from McWhan {\it et. al.}\cite{McWhan74} and Longo
 {\it et. al.}\cite{Longo70} respectively.

%-------------------------------------
\begin{center}
\begin{figure}[t]
  \includegraphics[width=7.3 cm]{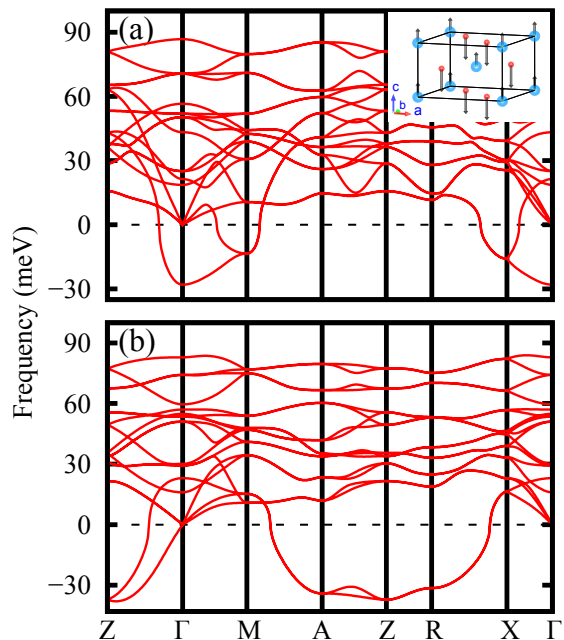}
  \caption{(Color online)
 (a) The phonon dispersion curves of \textit{R}-VO$_{2}$
 in the DFT. The inset figure shows the normal mode of softened phonon
 at $\Gamma$.
  (b) The phonon dispersion curves of \textit{R}-VO$_{2}$ in the DFT+$U$.
 Both figures show the negative phonon frequencies (imaginary frequencies)
 corresponding to the phonon softening instability. 
}\label{rph}
\end{figure}
\end{center}
%-------------------------------------

We have first obtained the electronic structures of 
\textit{R}-VO$_{2}$ in the DFT and the DFT+$U$.
In the DFT, stable metallic phases are obtained both in
the nonmagnetic and the magnetic band structure calculation,
in agreement with the experiment.
However, in the DFT+$U$, more stable insulating phase is obtained in
the magnetic band structure calculation,\cite{Mag}
which is seemingly in disagreement with the experiment.
In fact, the insulating phase of \textit{R}-VO$_{2}$ in the DFT+$U$
had been reported before.\cite{Korotin02,Liu10}
This discrepancy can be resolved by considering the competition 
between the magnetic instability and the structural instability 
in \textit{R}-VO$_{2}$.
Of course, in nature, the structural instability wins 
over the magnetic instability, and so 
\textit{R}-VO$_{2}$ undergoes the structural 
transition to \textit{M$_{1}$}-VO$_{2}$ rather than the magnetic transition.

Figure~\ref{rph} shows the phonon dispersion curves of \textit{R}-VO$_{2}$ 
in the DFT and DFT+$U$.
Four Raman modes (B$_{1g}$, E$_{g}$, A$_{1g}$, and 
B$_{2g}$) are obtained in both cases,
in agreement with experiments.\cite{Srivastava71,Schilbe02,Schilbe04}
The phonon softening instabilities are  
obtained both in the DFT (Fig.~\ref{rph}(a)) 
and in the DFT+$U$ (Fig.~\ref{rph}(b)), 
which imply that \textit{R}-VO$_{2}$ is not a stable structure. 
These results are expected
because \textit{R}-VO$_{2}$ is stable only at high temperature. 
The most noteworthy is the marked difference 
in the phonon dispersion curves
between the DFT and the DFT+$U$.  
In the DFT, the phonon softenings occur at 
{\bf q} = $\Gamma$, M and X (Fig.~\ref{rph}(a)),
while, in the DFT+$U$, they occur at {\bf q} = R, A and Z (Fig.~\ref{rph}(b)).
Indeed, 
the phonon softening at {\bf q} = R was once predicted to be responsible
for the transformation from \textit{R}-VO$_{2}$ to 
\textit{M$_{1}$}-VO$_{2}$.\cite{Brews70,Hearnt72,McWhan74,Terauchi78,Gervais85}

%-------------------------------------
\begin{center}
\begin{figure}[b]
  \includegraphics[width=8 cm]{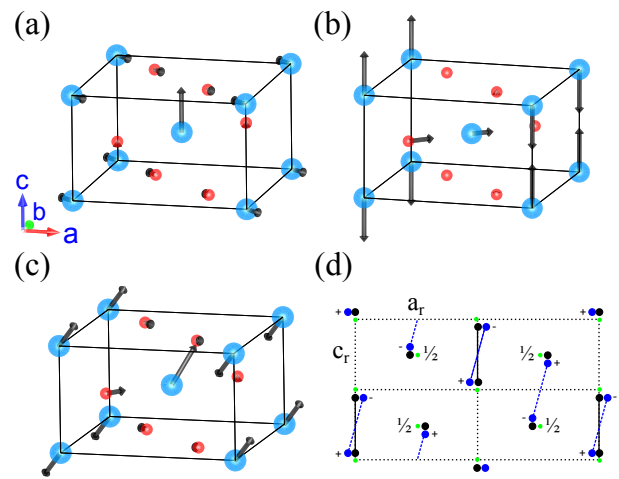}
  \caption{(Color online)
  The normal modes of the softened phonons of \textit{R}-VO$_{2}$ 
 in the DFT+$U$  at {\bf q} = R. There are two degenerate modes,
as shown in (a) and (b).
 (c) A linearly superposed mode at {\bf q} = R using two degenerate 
	modes of (a) and (b). 
  The blue and red balls represent V and O, respectively.
 (d) Schematic picture of 
	V-V pairing and distortion in \textit{M$_{1}$}-VO$_{2}$.
Blue, green, and black dots correspond to V ions in {M$_{1}$}-VO$_{2}$,
\textit{R}-VO$_{2}$, and {M$_{2}$}-VO$_{2}$), respectively.
}
\label{soft1}
\end{figure}
\end{center}
%-------------------------------------

We have examined the normal modes of the softened phonons. 
The normal mode at {\bf q} = $\Gamma$ in the DFT is
plotted in the inset of Fig.~\ref{rph}(a),
and those at {\bf q} = R in the DFT+$U$ are plotted 
in Fig.~\ref{soft1}(a) and (b).
Recall that the main structural changes from \textit{R}-VO$_{2}$ to 
\textit{M$_{1}$}-VO$_{2}$ are the dimerization 
and zig-zag distortion of V ions, as shown in Fig.~\ref{soft1}(d). 
It is evident that those lattice distortions cannot be
described by the normal mode at $\Gamma$ in the DFT.
The softened modes at M and X in the DFT do not describe the
structural distortions either.
In contrast, the normal modes at R in the DFT+$U$
are consistent with the lattice distortions of VO$_{2}$.
As shown in Fig.~\ref{soft1}(a) and (b), there are two degenerate 
softened phonon modes at R.
The first mode in Fig.~\ref{soft1}(a) represents the dimerizations of
half of V ions and the orthogonal displacements of the other half.
The second mode in Fig.~\ref{soft1}(b) is just the reverse of the first one.
Note that the mode predicted by Gervais 
{\it et al}.\cite{Gervais85} is close to the first mode.
Also similar modes to above two were once obtained by 
using a simple interatomic potential model.\cite{Woodley08} 
A linearly superposed mode using these two normal modes in Fig.~\ref{soft1}(c)
reveals simultaneous dimerizations and zig-zag distortions of V ions, 
in good agreement with the observed lattice distortions 
in \textit{M$_{1}$}-VO$_{2}$ (Fig.~\ref{soft1}(d)).
The other superposed mode also reveals similar displacements of V ions.
This agreement demonstrates that the softened mode at R in the DFT+$U$
describes the structural transition of VO$_{2}$ properly. 
The softened modes at A and Z in the DFT+$U$
are also related to the dimerizations of V ions,
but not directly to the zig-zag distortions of V ions.

%-------------------------------------
\begin{center}
\begin{figure}[t]
  \includegraphics[width=8.5 cm]{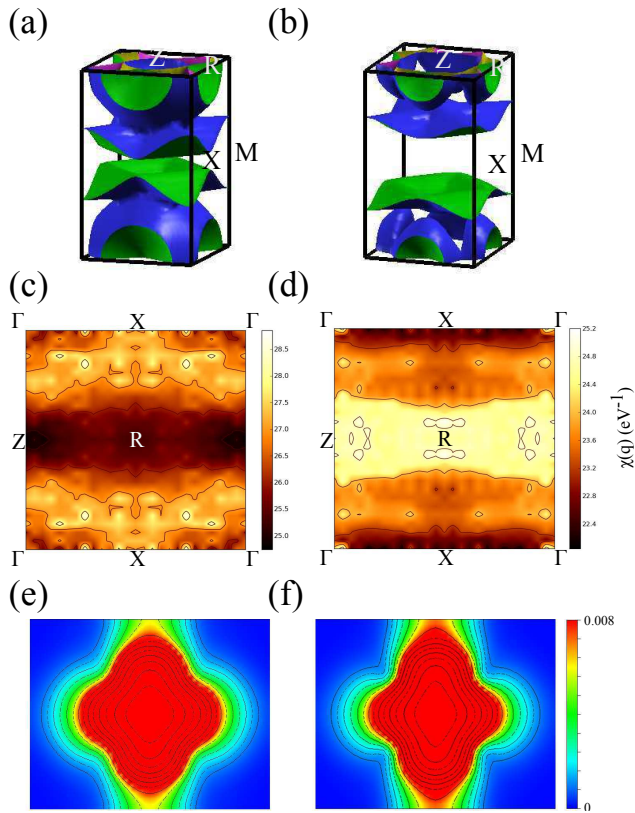}
  \caption{(Color online)
 (a),(b) The Fermi surface of \textit{R}-VO$_{2}$ 
	in the DFT and in the DFT+$U$, respectively.
 (c),(d) $\chi$({\bf q}) of \textit{R}-VO$_{2}$ 
	in the DFT and in the DFT+$U$, respectively.
 (e),(f) Local charge density near $E_F$ around V ion  
	(integrated over $\sim -1.2$ eV to $E_F$)
	in the DFT and in the DFT+$U$, respectively,
	in unit of (e/\AA$^{3}$).
}
\label{Fermi}
\end{figure}
\end{center}
%-------------------------------------  

The role of the Coulomb correlation effect in the 
structural transition in VO$_{2}$ has not 
been invoked as an essential factor before. The present phonon study 
clearly demonstrates that the Coulomb correlation in \textit{R}-VO$_{2}$ 
facilitates the Peierls-type structural transition.
Namely, in \textit{R}-VO$_{2}$, the Coulomb correlation effect 
and the Peierls distortion are mutually cooperating 
in driving the MIT and the structural transition.
To examine the $U$ effect in more detail,
we have checked the phonon dispersion curves with varying $U$.
With increasing $U$, we have found that
the softenings at {\bf q} = $\Gamma$, M and X disappear,
whereas the softenings at {\bf q} =  A, R and Z emerge
for $U \geq 4.0$ eV.
This feature indicates that large enough $U$ is required
to produce the right Peierls distortions in VO$_{2}$.

%-------------------------------------
\begin{center}
\begin{figure}[t]
  \includegraphics[width=8.0 cm]{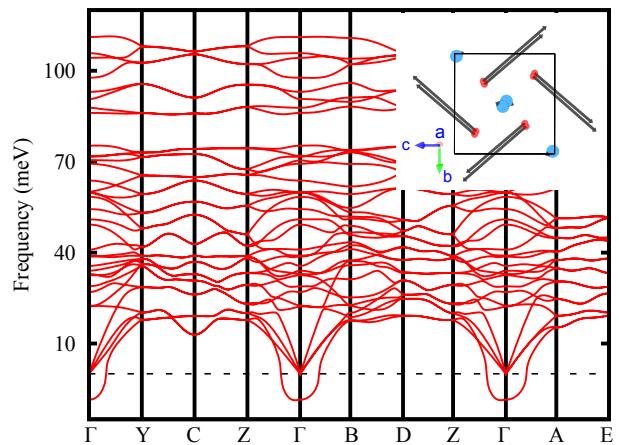}
  \caption{(Color online)
   The phonon dispersion curve of \textit{M$_{1}$}-VO$_{2}$ 
	at the pressure of 35.1 GPa
	in the DFT. 
   The inset shows 
  the normal mode of the softened phonon at {\bf q} = $\Gamma$.
The blue and red balls represent V and O, respectively.
}\label{m1soft}
\end{figure}
\end{center}
%-------------------------------------  

Figure~\ref{Fermi} shows the Fermi surface, the electric susceptibility
$\chi$({\bf q}), and the local charge density of \textit{R}-VO$_{2}$ 
in the DFT and the DFT+$U$ calculations.
By comparing the Fermi surfaces in Fig.~\ref{Fermi}(a) and (b),
one can notice that the Fermi surface becomes flatter in the DFT+$U$,
suggesting a possible nesting effect along the $c$-axis.\cite{Gupta77}
This change is more apparent in $\chi$({\bf q}).
As compared to $\chi$({\bf q}) in the DFT,
$\chi$({\bf q}) in the DFT+$U$ exhibits high values at {\bf q}=($x$,0,0.5)
(0 $\leq x \leq$ 0.5), which explains the origin of
phonon softenings at {\bf q}=($x$,0,0.5),
that is, at Z and R in Fig.~\ref{rph}(b).
Local charge density plots in Fig.\ref{Fermi} (e) and (f) 
also support the proper description of
the DFT+$U$ for the structural transition of \textit{R}-VO$_{2}$. 
As compared to the case of the DFT,
the charge density in the DFT+$U$ is seen to be more anisotropic, 
reflecting that more charges are  
accumulated in the bonding region along the $c$-axis.
This phenomenon arises from the correlation-induced orbital redistribution,
whereby $a_{1g}$ state is more occupied than other $e_g^{\pi}$ states.
Accordingly, the system becomes more one dimensional-like along the $c$-axis,
and so more susceptible to the Peierls transition.

Now we have investigated the stability of \textit{M$_{1}$}-VO$_{2}$
under pressure.
In the DFT calculation for \textit{M$_{1}$}-VO$_{2}$, 
we have obtained the insulating phase at the ambient pressure,
even though the energy gap is negligibly small ($\sim$ 0.03 eV).
This result is different from existing DFT results,
in which metallic phases were obtained 
for \textit{M$_{1}$}-VO$_{2}$.\cite{Wentzcovitch94,Eyert02,Liebsch05} 
This difference is thought to come from the full relaxation of volume and
ionic positions in our band calculations, which was not 
taken into account in the previous calculations. 
Without the relaxation, we also get the metallic phase for
\textit{M$_{1}$}-VO$_{2}$ in the DFT.  
In the DFT+$U$, we have obtained the insulating phase with energy gap of
$\sim$ 0.67 eV, which agrees well with the experimental gap of 
0.6-0.7 eV.\cite{Wentzcovitch94,Shin90} 

We have calculated the phonon dispersions of \textit{M$_{1}$}-VO$_{2}$
both in the DFT and the DFT+U.
In both cases, there are no phonon softening instabilities, 
which indicate that \textit{M$_{1}$}-VO$_{2}$ is a stable structure 
at the ambient pressure.
Interesting feature is obtained in the phonon dispersion  
of \textit{M$_{1}$}-VO$_{2}$ under pressure.
Figure~\ref{m1soft} shows the phonon dispersion curve 
of \textit{M$_{1}$}-VO$_{2}$ at 35.1 GPa in the DFT,
which manifests the phonon softening instability at $\Gamma$.
This softened mode implies that
\textit{M$_{1}$}-VO$_{2}$ would undergo a structural transition 
at some pressure $P \leq 35.1$ GPa.  
This result is indeed consistent with the   
experiments,\cite{Arcangeletti07,Marini08,Mitrano12}
in which the pressure-induced transition 
from insulating \textit{M$_{1}$}-VO$_{2}$
to a metallic phase of \textit{M$_{x}$}-VO$_{2}$ was observed.

The normal mode of the softened phonon at $\Gamma$ is 
depicted in the inset of Fig.~\ref{m1soft},
which shows the rotation of oxygen (O) octahedron with
slight displacements of V ions.
The transformation from \textit{M$_{1}$}-VO$_{2}$
to \textit{M$_{x}$}-VO$_{2}$ is expected to arise from
this rotation mode.
In the experiment under pressure, the anisotropic compression was observed 
in \textit{M$_{x}$}-VO$_{2}$.\cite{Mitrano12} 
Lattice constants $b$ and $c$ become softened and hardened,
respectively, while $a$ decreases regularly.
This anisotropy is reminiscent of the case of CrO$_{2}$,
which also has a pressure-induced structural transition from rutile 
to CaCl$_{2}$-type with the softened normal mode 
of the O octahedron rotation.
Anisotropic compression is also present in
CrO$_{2}$ of CaCl$_{2}$-type.\cite{Maddox06,Kim12}
This similarity suggests that \textit{M$_{x}$}-VO$_{2}$ 
would have a monoclinic structure of CaCl$_{2}$-type.

In conclusion,
based on the phonon dispersion studies in the DFT and the DFT+$U$,
we have demonstrated that 
the driving mechanism of the MIT and the structural transition in VO$_{2}$
is the Mott-Peierls transition. 
The Coulomb correlation effect in \textit{R}-VO$_{2}$ 
plays an essential role of assisting the Peierls transition.
The softened phonon mode at {\bf q} = R in the DFT+$U$  
describes properly the  structural transition
from \textit{R}-VO$_{2}$ to \textit{M$_{1}$}-VO$_{2}$.
Further, we have found that \textit{M$_{1}$}-VO$_{2}$ becomes 
unstable under high pressure due to the phonon softening at $\Gamma$. 
This softened mode is expected to 
have the relation to the pressure-induced transition 
from \textit{M$_{1}$}-VO$_{2}$ to \textit{M$_{x}$}-VO$_{2}$,
which is predicted to have a monoclinic structure of CaCl$_{2}$-type.

This work was supported by the NRF (No. 2009-0079947, No. 2011-0025237) 
and the KISTI supercomputing center (No. KSC-2012-C2-18).
S.K. acknowledges the support from the NRF project 
of Global Ph.D. Fellowship (No. 2011-0002351).


\begin{thebibliography}{99}
\bibitem{Mag}{The energy gaps and the magnetizations were obtained to be
	0.86 eV and $\sim 1.1 \mu_{B}$ for the ferromagnetic case,
	while 1.25 eV and $\sim  1.1 \mu_{B}$ for the antiferromagnetic case.}
\end{thebibliography}
\end{document}